\documentclass[oneside,12pt]{article}
\begin{document}

\title{\bf{
A Fresh Look at Entropy and the Second Law of Thermodynamics}}
\author{\vspace{5pt} Elliott H.~Lieb$^1$, and Jakob 
Yngvason$^{2}$\\ 
\vspace{-4pt}\small{$1.$ Departments of Physics and Mathematics, Jadwin 
Hall,} \\
\small{Princeton University, P.~O.~Box 708, Princeton, New Jersey
  08544}\\
\vspace{-4pt}\small{$2.$ Institut f\"ur Theoretische Physik, Universit\"at 
Wien}\\
\small{Boltzmanngasse 5, A 1090 Vienna, Austria}}
\date{}
\maketitle

\begin{abstract}The existence of entropy and its increase can be
understood without reference to either statistical mechanics or heat
engines.
\end{abstract}
\footnotetext{To be published in PHYSICS TODAY, April 2000. The publication
has 4 figures, which are not available in electronic form.
}

In days long gone, the second law of thermodynamics (which predated 
the first law) was regarded as perhaps the most perfect and 
unassailable law in physics.  It was even supposed to have 
philosophical import: It has been hailed for providing a proof of the 
existence of God (who started the universe off in a state of low 
entropy, from which it is constantly degenerating); conversely, it has 
been rejected as being incompatible with dialectical materialism and 
the perfectability of the human condition.

Alas, physicists themselves eventually demoted it to a lesser position 
in the pantheon---because (or so it was declared) it is ``merely'' 
statistics applied to the mechanics of large numbers of atoms.  
Willard Gibbs wrote: ``The laws of thermodynamics may easily be 
obtained from the principles of statistical mechanics, of which they 
are the incomplete expression'' [1]---and Ludwig Boltzmann expressed 
similar sentiments.

Is this really so?  Is it really true that the second law is merely an 
``expression'' of microscopic models or could it exist in a world that 
was featureless at the $10^{-8}$ cm level?  We know that statistical 
mechanics is a powerful tool for understanding physical phenomena and 
calculating many quantities, especially in systems at or near 
equilibrium.  We use it to calculate entropy, specific and latent 
heats, phase transition properties, transport coefficients and so on, 
often with good accuracy.  Important examples abound, such as Max 
Planck's realization that by staring into a furnace he could find 
Avogadro's number or Linus Pauling's highly accurate 
back-of-the-envelope calculation of the residual entropy of ice.  But 
is statistical mechanics essential for the second law?

In any event, it is still beyond anyone's computational ability 
(except in idealized situations) to account for a very precise, 
essentially infinitely accurate law of physics from statistical 
mechanical principles.  No exception has ever been found to the second 
law of thermodynamics---not even a tiny one.  Like conservation of 
energy (the ``first'' law) the existence of a law so precise and so 
independent of details of models must have a logical foundation that 
is independent of the fact that matter is composed of interacting 
particles.  Our aim here is to explore that foundation.  The full 
details can be found in [2].

As Albert Einstein put it, ``A theory is the more impressive the 
greater the simplicity of its premises is, the more different kinds of 
things it relates, and the more extended is its area of applicability.  
Therefore the deep impression which classical thermodynamics made upon 
me.  It is the only physical theory of universal content concerning 
which I am convinced that, within the framework of the applicability 
of its basic concepts, it will never be overthrown'' [3].

In an attempt to reaffirm the Second Law as a pillar of physics in its 
own right, we have returned to a little noticed movement that began in 
the 1950's with the work of Peter Landsberg [4], Hans Buchdahl [5], 
Gottfried Falk, Herbert Jung [6], and others (see [2] for references) 
and culminated in the book of Robin Giles [7], which must be counted 
one of the truly great, but unsung works in theoretical physics.  It 
is in these works that the concept of ``comparison'' (explained below) 
emerges as one of the key underpinnings of the second law.  The 
approach of these authors is quite different from lines of thought in 
the tradition of Sadi Carnot that base thermodynamics on the 
efficiency of heat engines. (See [8], for example,  for a modern exposition of 
the latter.)

\section*{The basic question}

The paradigmatic event that the second law deals with can be described 
as follows.  Take a macroscopic system in an  equilibrium state 
$X$ and place it in a room, together with a gorilla equipped with 
arbitrarily complicated machinery (a metaphor for the rest of the 
universe), and a weight---and close the door.  As in the old 
advertisement for indestructible luggage, the gorilla can do anything 
to the system---including tearing it apart.  At the end of the day, 
however, the door is opened and the system is found in some other 
equilibrium state, $Y$, the gorilla and machinery are found in their 
original state, and the only other thing that has possibly changed is 
that the weight has been raised or lowered.  Let us emphasize that 
although our focus is on equilibrium states, the processes that take 
one such state into another can be arbitrarily violent.  The gorilla 
knows no limits.

The question that the second law answers is this: What distinguishes 
those states $Y$ that can be reached from $X$ in this manner from 
those that cannot?  The answer: There is a function of the equilibrium 
states, called entropy and denoted by $S$, that characterizes 
the possible pairs of equilibrium states $X$ and $Y$ by the inequality 
$S(X)\leq S(Y)$.  The function can be chosen to be additive (in a 
sense explained below), and with this requirement it is unique, 
up to a change of scale.  Our main point is that the existence of 
entropy relies only on a few basic principles, independent of any 
statistical model---or even of atoms.

What is exciting about this apparently innocuous statement is the 
uniqueness of entropy, for it means that all the different methods for 
measuring or computing entropy must give the same answer.  The usual 
textbook derivation of entropy as a state function, starting with some 
version of ``the second law'', proceeds by considering certain slow, 
almost reversible processes (along adiabats and isotherms).  It is not 
at all evident that a function obtained in this way can contain any 
information about processes that are far from being slow or 
reversible.  The clever physicist might think that with the aid of 
modern computers, sophisticated feedback mechanisms, unlimited amounts 
of mechanical energy (represented by the weight) and lots of plain 
common sense and funding, the system could be made to go from an 
equilibrium state $X$ to $Y$ that could not be achieved by the 
primitive quasistatic processes used to define entropy in the first 
place.  This cannot happen, however, no matter how clever the 
experimentalist or how far from equilibrium one travels!

What logic lies behind this law?  Why can't one gorilla undo what 
another one has wrought?  The atomistic foundation of this logic is 
not as simple as is often suggested.  It not only concerns things like 
the enormous number of atoms involved ($10^{23}$), but also other 
aspects of statistical mechanics that are beyond our present 
mathematical abilities.  In particular, the interaction of a system 
with the external world (represented by the gorilla and machinery) 
cannot be described in any obvious way by Hamiltonian mechanics.  
Although irreversibility is an important open problem in statistical 
mechanics, it is fortunate that the logic of thermodynamics itself is 
independent of atoms and can be understood without knowing its source.

The founders of thermodynamics---Rudolf Clausius, Lord Kelvin, Max 
Planck, Constantin Carath\'eodory, and so on---clearly had 
transitions between equilibrium states in mind when they stated the 
law in sentences such as ``No process is possible, the sole result of 
which is that a body is cooled and work is done'' (Kelvin).  Later it 
became tacitly understood that the law implies a continuous increase 
in some property called entropy, which was supposedly defined for 
systems out of equilibrium.  The ongoing, unsatisfactory debates (see 
referce [9, for example) about the definition of this nonequilibrium 
entropy and whether it increases shows, in fact, that what is 
supposedly ``easily'' understood needs clarification.  Once again, it 
is a good idea to try to understand first the meaning of entropy for 
equilibrium states---the quantity that our textbooks talk about when 
they draw Carnot cycles.  In this article we restrict our attention to 
just those states; by ``state'' we always mean ``equilibrium state''.  
Entropy, as the founders of thermodynamics understood the quantity is 
subtle enough, and it is worthwhile to understand the ``second law'' 
in this restricted context.  To do so it is not necessary to decide 
whether Boltzmann or Gibbs had the right view on irreversibility.  
(Their views are described in Joel L. Lebowitz's article ``Boltzmann's 
Entropy and Time's Arrow'',  Physics Today, September 1993, page 32.)

\section*{The basic concepts}

To begin at the beginning, we suppose we know what is meant by a 
thermodynamic system and equilibrium states of such a system.  
Admittedly these are not always easy to define, and there are 
certainly systems, such as a mixture of hydrogen and oxygen or an 
interstellar ionized gas, that can behave as if they are in 
equilibrium even if it is not truly so.  The prototypical system is a 
``simple system'', consisting of a substance in a container with a 
piston.  But a simple system can be much more complicated than that.  
Besides its volume it can have have other coordinates, which can be 
changed by mechanical or electrical means---shear in a solid, or 
magnetization, for example.  In any event, a state of a simple system 
is described by a special coordinate $U$, which is its energy, and one 
or more other coordinates (such as the volume $V$) called work 
coordinates.  An essential point is that the concept of energy, which 
we know about from the moving weight and Newtonian mechanics, can be 
defined for thermodynamic systems.  This fact is the content of the 
first law of thermodynamics.

Another type of system is a ``compound system'', which consists of 
several different or identical independent, simple systems.  By means 
of mixing or chemical reactions, systems can be created or destroyed.

Let us briefly discuss some concepts that are relevant for systems and 
their states, which are denoted by capital letters such as 
$X,X',Y,\ldots$.  Operationally, the composition, denoted $(X,X')$, of 
two states $X$ and $X'$ is obtained simply by putting one system in a 
state $X$ and one in a state $X'$ side by side on the experimental 
table and regarding them jointly as a state of a new, compound system.  
For instance, $X$ could be a glass containing 100 g of whiskey at 
standard pressure and $20^\circ$ C, and $X'$ a glass containing 50 g 
of ice at standard pressure and $0^\circ$ C. To picture $(X,X')$ one 
should think the two glasses standing on a table without touching each 
other. 

Another operation is the ``scaling'' of a state $X$ by a factor 
$\lambda>0$, leading to a state denoted $\lambda X$.  Extensive 
properties like mass, energy and volume are multiplied by $\lambda$, 
while intensive properties such as pressure stay intact.  For the 
states $X$ and $X'$ as in the example above the example above ${1\over 
2} X$ is 50 g of whiskey at standard pressure and $20^\circ$ C, and 
${1\over 5}X'$ is 10 g of ice at standard pressure and $0^\circ$ C. 
Compound systems scale in the same way: ${1\over 5}(X,X')$ is 20 g of 
whiskey and 10 g of ice in separate glasses with pressure and 
temperatures as before.

A central notion is adiabatic accessibility.  If our gorilla can take 
a system from $X$ to $Y$, as described above---that is, if the only 
net effect of the action, besides the state change of the system, is 
that a weight has possibly been raised or lowered, we say that $Y$ is 
adiabatically accessible from $X$ and write $X\prec Y$ (the symbol 
$\prec$ is pronounced ``precedes'').  It has to be emphasized that for 
macroscopic systems this relation is an absolute one: If a transition 
from $X$ to $Y$ is possible at one time, then it is {\it always} 
possible (that is, it is reproducible), and if it is impossible at one 
time it {\it never} happens.  This absolutism is guaranteed by the 
large powers of 10 involved---the impossibility of a chair's 
spontaneous jumping up from the floor is an example.

\section*{ The role of entropy}

Now imagine that we are given a list of all possible pairs of states 
$X,Y$ such that $X\prec Y$.  The foundation on which thermodynamics 
rests, and the essence of the second law, is that this list can be 
simply encoded in an {entropy function $S$ on the set of all states of 
all systems (including compound systems) so that when $X$ and $Y$ are 
related at all, then
$$ 
X\prec Y \  \ \hbox{\rm if and only if}\  \  S(X) \leq S(Y) \ .  
$$
Moreover, the entropy function  can be chosen in such a way that 
if $X$ and $X'$ are states of two  (different or identical) 
systems,  then the entropy of the compound system in this pair of states
is given by 
$$ S(X,X') = S(X) + S(X').  
$$
This additivity of
entropy is a highly nontrivial assertion. Indeed, it is one of the
most far reaching properties of the second law. In compound systems
such as the whiskey/ice example above, all states $(Y,Y')$ such that
$X\prec Y$ and $X'\prec Y'$ are adiabatically accessible from $(X,X')$.
For instance, by letting a falling weight run an electric generator one
can stir the whiskey and also melt some ice.  But it is important to
note that $(Y,Y')$ can be adiabatically accessible from $(X,X')$
without $Y$ being adiabatically accessible from $X$. Bringing the two
glasses into contact and separating them again is adiabatic for the
compound system but the resulting cooling of the whiskey is not 
adiabatic for the whiskey alone.  The fact that the inequality
$S(X)+S(X')\leq S(Y)+S(Y')$ {\it exactly} characterizes the possible
adiabatic transitions for the compound system, even when $S(X)\geq
S(Y)$, is quite remarkable. It means that it is sufficient to know the
entropy of each part of a compound system in order to decide which
transitions due to interactions between these parts  (brought about
by the gorilla) are possible.

Closely related to additivity is extensivity, or scaling of entropy, 
$$S(\lambda X)=\lambda S(X),$$ which means that the entropy of an 
arbitrary mass of a substance is determined by the entropy of some 
standard reference mass, such as 1 kg of the substance.  Without this 
property engineers would have to use different steam tables each time 
they designed a new engine.

In traditional presentations of thermodynamics, based for example on 
Kelvin's principle given above, entropy is arrived at in a rather 
roundabout way which tends to obscure its connection with the relation 
$\prec$.  The basic message we wish to convey is that existence and 
uniqueness of entropy are equivalent to certain simple properties of 
the relation $\prec$.  This equivalence is the concern of [2].

An analogy leaps to mind: When can a vector-field, ${\bf E}(x)$, be 
encoded in an ordinary function (potential), $\phi(x)$, whose gradient 
is ${\bf E}$?  The well-known answer is that a necessary and 
sufficient condition is that ${\rm curl}\, {\bf E} =0$.  The 
importance of this encoding does not have to be emphasized to 
physicists; entropy's role is similar to the potential's role and the 
existence and meaning of entropy are not based on any formula such as 
$S=-\Sigma_i p_i\ln p_i$, involving probabilities $p_i$ of 
``microstates''.  Entropy is derived (uniquely, we hope) from the list 
of pairs $X\prec Y$; our aim is to figure out what properties of this 
list (analogous to the curl-free condition) will allow it to be 
described by an entropy.  That entropy will then be endowed with an 
unambiguous physical meaning independent of anyone's assumptions about 
``the arrow of time'', ``coarse graining'' and so on.  Only the list, 
which is given by physics, is important for us now.

The required properties of $\prec$ do {\it not} involve concepts like 
``heat'' or ``reversible engines'', not even ``hot'' and ``cold'' are 
needed.  Besides the ``obvious'' conditions ``$X\prec X$ for all $X$'' 
(reflexivity) and ``$X\prec Y$ and $Y\prec Z$ implies $X\prec Z$'' 
(transitivity) one needs to know that the relation behaves reasonably 
with respect to the composition and scaling of states.  By this we 
mean the following:

\begin{itemize} \item[{$\bullet$}] Adiabatic accessibility is 
consistent with the composition of states: $X\prec Y$ and $Z\prec W$ 
implies $(X,Z)\prec (Y,W)$.  \item[{$\bullet$}] Scaling of states does not 
affect adiabatic accessibility: If $X\prec Y$, then $\lambda X\prec 
\lambda Y$.  \item[{$\bullet$}] Systems can be cut adiabatically into two 
parts: If $0<\lambda<1$, then $X\prec ((1-\lambda)X,\lambda X)$, and 
the recombination of the parts is also adiabatic: 
$((1-\lambda)X,\lambda X)\prec X$.  \item[{$\bullet$}] Adiabatic accessibility 
is stable with respect to small perturbations: If $(X,\varepsilon 
Z)\prec (Y, \varepsilon W)$ for arbitrarily small $\varepsilon>0$, 
then $X\prec Y$. \end{itemize}

These requirements are all very natural.  In fact, in traditional 
approaches they are usually taken for granted, without mention.  They 
are not quite sufficient, however, to define entropy.  A crucial 
additional ingredient is the comparison hypothesis for the relation 
$\prec$.  In essence, this is the hypothesis that all equilibrium 
states, simple or compound, can be grouped into classes, such that if 
$X$ and $Y$ are in the same class, then either $X\prec Y$ or $Y\prec 
X$.  In nature, a class consists of all states with the same mass and 
chemical composition---that is, with the same amount of each of the 
chemical elements.  If chemical reactions and mixing processes are 
excluded, the classes are smaller and may be identified with the 
``systems'' in the usual parlance.  But it should be noted that 
systems can be compound, or consist of two or more vessels of 
different substances.  In any case, the role of the comparison 
hypothesis is to insure that the list of pairs $X\prec Y$ is 
sufficiently long.  Indeed, we shall give an example later where the 
list of pairs satisfies all the other axioms, but which is {\it not} 
describable by an entropy function.

\section*{The construction of entropy}

Our main conclusion (which we do not claim isobvious, but whose proof 
can be found in reference [2]) is that the existence and uniqueness of 
entropy is a consequence of the comparison hypothesis and the 
assumptions about adiabatic accessibility stated above.  In fact, if 
$X_0$, $X$ and $X_1$ are three states of a system and $\lambda$ is any 
scaling factor between 0 and 1, then either $X\prec ((1-\lambda) 
X_0,\lambda X_1)$ or $((1-\lambda) X_0,\lambda X_1)\prec X$ must be 
true, by the comparison hypothesis.  If {\it both} alternatives hold, 
then the properties of entropy demand that
$$
S(X)=(1-\lambda) S(X_0)+\lambda S(X_1).
$$
If $S(X_0)\neq S(X_1)$ this equality can hold for at most one
$\lambda$. With $X_0$ and $X_1$ as reference states, the entropy is
therefore {\it fixed}, apart from two free constants, namely the values
$S(X_0)$ and $S(X_1)$.

{}From the properties of the relation $\prec$ listed above, one can 
show that there is, indeed, always a $0\leq \lambda\leq 1$ with the 
required properties, provided that $X_0\prec X\prec X_1$.  It is the 
{\it largest} $\lambda$, denoted $\lambda_{\rm max}$, such that 
$((1-\lambda) X_0,\lambda X_1)\prec X$.  Defining the entropies of the 
reference states arbitrarily as $S(X_0)=0$ and $S(X_1)=1$ unit, we 
obtain the following simple {\it formula for entropy}:
$$
S(X)=\lambda_{\rm max}\ \hbox{\rm units}.
$$
The scaling factors $(1-\lambda)$ and $\lambda$ measure the amount of 
substance in the states $X_0$ and $X_1$ respectively.  The formula for 
entropy can therefore be stated in the following words: $S(X)$ is the 
maximal fraction of substance in the state $X_{1}$ that can be 
transformed adiabatically} (that is, in the sense of $\prec$) into the 
state $X$ with the aid of a complementary fraction of substance in the 
state $X_{0}$.  This way of measuring $S$ in terms of substance is 
reminiscent of an old idea, suggested by Pierre Laplace and Antoine 
Lavoisier, that heat be measured in terms of the amount of ice melted 
in a process.  As a concrete example, let us assume that $X$ is a 
state of liquid water, $X_{0}$ of ice and $X_{1}$ of vapor.  Then 
$S(X)$ for a kilogram of liquid, measured with the entropy of a 
kilogram of water vapor as a unit, is the maximal fraction of a 
kilogram of vapor that can be transformed adiabatically into liquid in 
state $X$ with the aid of a complementary fraction of a kilogram of 
ice.

In this example the maximal fraction $\lambda_{\rm max}$ cannot be 
achieved by simply exposing the ice to the vapor, causing the former 
to melt and the latter to condense.  This would be an irreversible 
process---that is, it would not be possible to reproduce the initial 
amounts of vapor of ice adiabatically (in the sense of the definition 
given earlier) from the liquid.  By contrast, $\lambda_{\rm max}$ is 
uniquely determined by the requirement that one can pass adiabatically 
from $X$ to $((1-\lambda_{\rm max})X_{0}, \lambda_{\rm max}X_{1})$ 
{\it and} vice versa.  For this transformation it is necessary to 
extract or add energy in the form of work---for example by running a 
little reversible Carnot machine that transfers energy between the 
high-temperature and low-temperature parts of the system.
We stress, however, that neither the concept of a ``reversible 
Carnot machine" nor that of ``temperature" is needed for the logic 
behind the formula for entropy given above.  We mention these concepts 
only to relate our definition of entropy to concepts for which the 
reader may have an intuitive feeling.

By interchanging the roles of the three states, the definition of 
entropy is easily extended to situations where $X\prec X_{0}$ or 
$X_{1}\prec X$.  Moreover, the reference points $X_{0}$ and $X_{1}$, 
where the entropy is defined to be 0 and 1 unit respectively, can be 
picked consistently for different systems such that the entropy will 
satisfy the crucial additivity and extensivity conditions
$$S(X,X')=S(X)+S(X')\qquad {\rm and}\qquad S(\lambda X)=\lambda S(X).$$

It is important to understand that once the existence and uniqueness 
of entropy has been established one need not rely on the $\lambda_{\rm 
max}$ formula displayed above to determine it in practice.  There are 
various experimental means to determine entropy that are usually much 
more practical.  The standard method consists of measuring pressures, 
volumes and temperatures (on some empirical scale), as well as 
specific and latent heats.  The empirical temperatures are converted 
into absolute temperatures $T$ (by means of formulas that follow from 
the mere existence of entropy but do not involve $S$ directly), and 
the entropy is computed by means of formulas like $\Delta S=\int 
(dU+PdV)/T$, with $P$ the pressure.  The existence and uniqueness of 
entropy implies that this formula is independent of the path of 
integration.

\section*{Comparability of states}

The possibility of defining entropy entirely in terms of the relation 
$\prec$ was first clearly stated by Giles [7].  (Giles's definition is 
different from ours, albeit similar in spirit.)  The importance of the 
comparison hypothesis had been realized earlier, however [4, 5, 6].  
All these authors take the comparison hypothesis as a  
postulate---that is, they do not attempt to justify it from other simpler premises.  
However, it is in fact possible to {\it derive} comparability for any 
pair of states of the same system from some natural and directly 
accessible properties of the relation $\prec$ [2].  In this derivation 
of comparison the customary parametrization of states in terms of 
energy and work coordinates is used, but it has to be stressed that 
such parametrizations are irrelevant, and therefore not used, for our
definition of entropy---once the comparison 
hypothesis is established.

To appreciate the significance of the comparison hypothesis it may be 
helpful to consider the following example.  Imagine a world whose 
thermodynamical sytems consist exclusively of incompressible solid 
bodies.  Moreover, all adiabatic state changes in this world are 
supposed to be obtained by means of the following elementary 
operations:

\begin{itemize}\item[{$\bullet$}]Mechanical rubbing of 
the individual systems, increasing their energy.  
\item[{$\bullet$}]Thermal equilibration in the conventional sense (by 
bringing the systems into contact.)\end{itemize}

The state space of the compound system consisting of two identical 
bodies, 1 and 2, can be paramertized by their energies, $U_{1}$ and 
$U_{2}$. If $X=(U_1,U_2)$ and $Y==(U_1',U_2')$ are such that 
$U_1'< U_1< U_2 < U_2'$ and $U_1+U_2 < U_1'+U_2'$ then one finds that
that neither $X\prec Y$ 
nor $Y\prec X$ holds.  The comparison hypothesis is therefore violated 
in this hypothetical example, and it is not possible to characterize 
adiabatic accessibility by means of an additive entropy function.  A 
major part of our work consists of understanding why such situations 
do not happen---why the comparison hypothesis appears to be true in 
the real world.

The derivation of the comparison hypothesis is based on an analysis of 
simple systems, which are the building blocks of thermodynamics.  As 
already mentioned the states of such systems are described by one 
energy coordinate $U$ and at least one work coordinate, like the 
volume $V$.  The following concepts play a key role in this analysis:

\begin{itemize} \item[{$\bullet$}] The possibility of 
forming ``convex combinations'' of states of simple systems with 
respect to the energy $U$ and volume $V$ (or other work coordinates).  
This means that given any two states $X$ and $Z$ of one kilogram of 
our system one can pick any state $Y$ on the line between them in 
$U$, $V$ space and, by taking appropriate fractions $\lambda$ and 
$1-\lambda$ in states $X$ and $Z$, respectively, there will be an 
adiabatic process taking this pair of states into the state $Y$. This 
process is usually quite elementary. For example, for gases and 
liquids one need only remove the barrier that separates the two 
fractions of the system. The fundamental property of entropy increase 
will then tell us that $S(Y)\geq \lambda S(X)+(1-\lambda)S(Z)$. As 
Gibbs emphasized, this ``concavity'' is the basis for thermodynamical 
stability---namely positivity of specific heats and 
compressibilities. \item[{$\bullet$}] The existence of 
at least one irreversible adiabatic state change, starting from 
any given state.  In conjuction with concavity of $S$ this seemingly 
weak requirement excludes the possibility that the entropy is constant 
in a whole neighborhood of some state.  The classical formulations of 
the second law follow from this.  \item[{$\bullet$}] The concept of  
thermal equilibrium between simple systems, which means, 
operationally, that no state changes takes place when the systems are 
allowed to exchange energy with each other at fixed work coordinates.  
The zeroth law of thermodynamic says that if two systems are in 
thermal equilibrium with a third, then they are in thermal 
equilibrium with one another.  This property is essential for the 
additivity of entropy, because it allows a consistent adjustment of 
the entropy unit for different systems.  It leads to a definition of 
temperature by the usual formula $1/T=(\partial S/\partial 
U)_{V}$.\end{itemize}

Using these notions (and a few others of a more technical nature) the
comparison hypothesis can be established for all simple systems and
their compounds.

It is more difficult to justify the comparability of states if mixing 
processes or chemical reactions are taken into account.  In fact, 
although a mixture of whiskey and water at $0^\circ$ C is obviously 
adiabatically accessible from separate whiskey and ice by pouring
whiskey from one glass onto the rocks in the other glass, it is not
possible to reverse this process adiabatically.  Hence it is not clear
that a block of a frozen whiskey/water mixture at $-10^\circ$ C, say,
is at all related in the sense of $\prec$ to a state in which whiskey and
water are in separate glasses.  Textbooks usually appeal here to {\it
gedanken} experiments with ``semipermeable
membrane'' that let only water molecules through and withhold the
whiskey molecules, but such membranes really exist only in the mind 
[10].
However, without invoking any such device, it turns out to be possible
to shift the entropy scales of the various substances in such a way
that $X\prec Y$ always implies $S(X)\leq S(Y)$. The converse assertion,
namely, $S(X)\leq S(Y)$ implies $X\prec Y$ provided $X$ and $Y$
have the same chemical composition, cannot be guaranteed {\it a priori} 
for mixing and chemical reactions, but
it is empirically testable and appears to be true in the real world.
This aspect of the second law, comparability, is not usually 
stressed, but it is important; it is challenging to figure out how to 
turn the frozen whiskey/water block into a glass of whiskey and a 
glass of water without otherwise changing the universe, except for 
moving a weight, but such an adiabatic process is possible.

\section*{What has been gained?}

The line of thought that started more than forty years ago has led to 
an axiomatic foundation for thermodynamics.  It is appropriate to ask 
what if anything has been gained compared to the usual approaches 
involving quasi-static processes and Carnot machines on the one hand 
and statistical mechanics on the other hand.  There are several 
points.  One is the elimination of intuitive, but hard-to-define 
concepts like ``hot'', ``cold'' and ``heat''.  Another is the 
recognition of entropy as a codification of possible state changes, 
$X\prec Y$, that can be accomplished without changing the rest of the 
universe in any way except for moving a weight.  Temperature is 
eliminated as an {\it a priori} concept and appears in its natural 
place as a quantity derived from entropy and whose consistent 
definition really depends on the existence of entropy, rather than the 
other way around.  To define enetropy, there is no need for special 
machines and processes on the empirical side, and there is no need for 
assumptions about models on the statistical mechanical side.  Just as 
energy conservation was eventually seen to be a consequence of time 
translation invariance, in like manner entropy can be seen to be a 
consequence of some simple properties of the list of state pairs 
related by adiabatic accessibility.

If the second law can be demystified, so much the better. If it can be
seen to be a consequence of simple, plausible notions then, as Einstein
said, it cannot be overthrown.

\section*{Acknowledgements} We are grateful to Shivaji Sondhi and Roderich Moessner for helpful
suggestions. Lieb's work was supported by NSF grant PHY 9820650. 
Yngvason's work was supported by 
the Adalsteinn Kristj\'ansson Foundation and the University of Iceland.

\section*{References}
\noindent
[1] C. Kittel and H. Kroemer, {\it Thermal Physics}, p. 57, Freeman, NY (1980).

\noindent
[2] E.H. Lieb and J. Yngvason, Physics Reports {\bf 310}, 1 (1999).

\noindent
[3] A. Einstein,  Autobiographical Notes in {\it Albert Einstein:
Philosopher-Scientist} P. A. Schilpp (ed.), Library of Living
Philosophers, vol VII, p. 33, Cambridge University Press, London, 1970.

\noindent
[4] P.T. Landsberg, Rev.~Mod.~Phys.\ {\bf 28}, 363 (1956).

\noindent
[5] H.A. Buchdahl, {\it The Concepts of Classical Thermodynamics}, 
Cambridge University Press, London (1966). 

\noindent
[6] G. Falk and H. Jung {\it Handbuch der Physik,} {\bf III/2},
S.~Fl\"ugge ed.,  p.~199 Springer, Berlin (1959).

\noindent
[7] R.Giles, {\it Mathematical Foundations of Thermodynamics,}
Pergamon, Oxford (1964).

\noindent
[8] D.R.Owen, {\it A First Course in the Mathematical Foundations of
Thermodynamics,} Springer, Heidelberg (1984).  J. Serrin, Arch. Rat. 
mech. Anal. {\bf 70}, 355 (1979).
M. Silhav\'y, {\it The
Mechanics and Thermodynamics of Continuous Media,} Springer,
Heidelberg (1997). C. A. truesdell, S. Baharata, {\it The Concepts 
and Logic of Thermodynamics as a Theory of heat Engines}, Heidelberg, 
(1977).

\noindent
[9] J.L. Lebowitz, I. Prigogine, and D.Ruelle, Physica A {\bf 263}, 
516, 528, 540 (1999).

\noindent
[10] E. Fermi, {\it Thermodynamics}, Dover, NY, (1956),  page 101.

\end{document}